\documentclass[twocolumn,showpacs,prl]{revtex4}
\usepackage{graphicx} 

\begin{document}
\title{Wall-Enhanced Convection in Vibrofluidized Granular Systems}
\author{J. Talbot$^{1,2}$ and P. Viot$^{2}$} \affiliation{
$^1$Department   of  Chemistry  and  Biochemistry,  Duquesne   University,
Pittsburgh,  PA  15282-1530, $^2$Laboratoire  de  Physique
Th{\'e}orique des  Liquides, Universit{\'e}  Pierre et  Marie Curie,  4, place
Jussieu,\\  75252 Paris   Cedex, 05  France}
 
\bibliographystyle{prsty} 
\begin{abstract}
An  event-driven  molecular   dynamics  simulation  of inelastic  hard
spheres  contained  in  a cylinder  and subject  to  strong  vibration
reproduces accurately  experimental  results\cite{WildmanHP01} for   a
system  of vibrofluidized glass  beads. In particular,  we are able to
obtain the  velocity field  and  the density  and temperature profiles
observed experimentally.  In addition,  we show that the appearance of
convection  rolls is    strongly   influenced by   the  value of   the
sidewall-particle restitution coefficient.  Suggestions for  observing
more complex convection patterns are proposed.
\end{abstract}
\maketitle
\section{Introduction}
If  sufficient power is supplied  to a granular  system contained in a
cylindrical container by a   vertically vibrating  wall,  fluidization
occurs. Once a steady  state  is obtained,  the  energy input  to  the
system by the wall  is, on average, balanced  by the energy dissipated
in inelastic grain-grain and grain-wall  collisions.  This energy flux
is  responsible for various  effects including  non-linear temperature
profiles,     heap     formation\cite{EvesqueR89}    and    convection
\cite{WildmanHP01,WildmanHHPA00}.  Recent experiments using   positron
emission  particle  tracking (PEPT)\cite{WildmanHP01}   have  observed
buoyancy-driven convection in a highly fluidized granular system for a
range   of  grain numbers  and  shaker  amplitudes.   While convection
effects  have   been observed   in   a two-dimensional  model   system
\cite{RRC00,G97}, no simulation studies
of  a three dimensional system corresponding   to the PEPT experiments
have yet been reported.

We study the  behavior  of a  model  system in  which the   grains are
modeled  as inelastic hard spheres.   Most  of the previous studies of
this model focused  on the homogeneous  cooling state, clustering, and
kinetic   and    hydrodynamic  theory     for particles  with    small
inelasticity\cite{van-NoijeE00,van-NoijeETP99,Javier-BreyDCS98}. Unlike
many  other many-body systems,  the model studied here closely matches
the  experimental system.  In particular,  the  number of particles is
identical and we   use values of particle-particle   and particle-wall
restitution   coefficients  that  were    determined  by   independent
experimental  measurements.    We   show that  the    model reproduces
accurately   the   experimentally  observed    density  and   granular
temperature     profiles\cite{WHP01}.    Moreover, for     a  range of
conditions,  a velocity field   exists  for  which   there is  a   net
circulation.   Interestingly,  the  direction  and  intensity   of the
circulation  are strong functions   of the  particle-wall  restitution
coefficient.  In addition, we suggest a modification to the experiment
that may generate new convection patterns.

\section{Model and simulation} 
The system  consists of a number  $N$  of hard spheres  contained in a
cylinder of radius $R$.   The spheres collide inelastically with  each
other and with the side walls with coefficients of restitution $c$ and
$c_w$, respectively. Between collisions, the  spheres are subject to a
downward constant  acceleration  due to the  (vertical)  gravitational
field.  Energy is  injected into the  system by the  bottom wall which
vibrates  with a   symmetric  saw-tooth profile  characterized  by  an
amplitude $A$ and a period $T$.  We do not expect  the behavior of the
system to be strongly dependent on the form of the profile\cite{MB97}.

The equations describing the  dynamics, which follow from conservation
of momentum,  for the  particle side-wall, particle bottom-wall  and
particle particle collisions are:
\begin{equation}
{\bf v}'_{i,r}={\bf v}_{i,r}-(1+c_w)({\bf v}_{i,r}.\hat{{\bf r}}_i)\hat{{\bf r}}_i
\end{equation}
\begin{equation}
v'_{i,z}=2v_{w}-v_{i,z}
\end{equation}
\begin{equation}
{\bf v}'_{i,j}={\bf v}_{i,j}\pm \frac{1+c}{2}[({\bf v}_j-{\bf
v}_i).\hat{{\bf n}}]\hat{{\bf n}}
\end{equation}
respectively where ${\bf v}_{i,r}$ is  the velocity of particle $i$ in
the $x,y$ plane, $v_{w}$ is the  velocity of the  bottom wall, $v_{i,z}$ is the
$z$  vertical component of the  velocity  of particle $i$,  $\hat{{\bf
r}}$ is the  unit position vector of particle  $i$ and $\hat{{\bf n}}$
is the unit center-to-center vector between the colliding pair $i$ and
$j$. Note that we have  taken  the tangential restitution  coefficient
for  sphere-sphere, and    sphere-wall, collisions equal   to
one.    The normal  restitution   coefficient  for  sphere-bottom wall
collisions was also taken as unity.

One  problem with the  event-driven  simulation is  that a  sphere may
collide more   and more frequently  with the  side wall  as the radial
component   of  its momentum  is  dissipated.  To  avoid the inelastic
collapse\cite{DeltourB97}, which has  been observed in simulations  of
bulk  granular systems, we simply  inject a small  amount of energy in
the radial direction  when the radial velocity  falls below a  cut-off
value. This condition arises infrequently and its handling in this way
has no discernible effect on the results.

\begin{figure}
\resizebox{7.5cm}{!}{\includegraphics{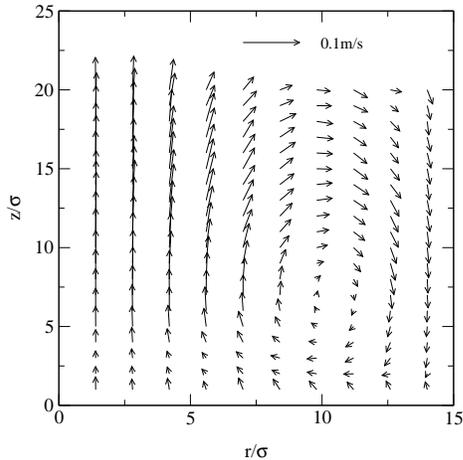}}
\caption{Mean velocity field for a  vibrofluidized system of inelastic hard
spheres in the $(r,z)$ plane. The wall-particle restitution coefficient is
$c_w=0.68$. Other parameters are given in the text }\label{fig:1}
\end{figure}

To reproduce the experimental conditions of Wildman et al
\cite{WildmanHP01} we used parameter values $N=1050$, $c=0.91$,
$c_w=0.68$. Taking  the unit of  length  as the  sphere diameter, $\sigma$, the
cylinder radius is $R/\sigma =14$. Since the base-wall particle coefficient is
not given,  we assume a  value of  one.  The shaker  amplitude is thus
left as the only    adjustable parameter for  the comparison   of  the
simulation with the experiments.  We  performed different runs with  a
dimensionless shaking amplitude varying  between $0.05$ and  $0.6$. To
generate the initial configuration, spheres were inserted sequentially
and randomly in the cylinder without overlap. A preliminary simulation
was  then performed, typically  for $5000$ collisions per particle, in
order  to  allow the  system  to reach  the stationary non-equilibrium
state.

Figure~\ref{fig:1} displays the mean velocity  field for the system of
inelastic spheres in the $(r,z)$ plane ($r$  denotes the radial distance
from the  axis  of the cylinder)   when the reduced shaker amplitude  is
$A/\sigma =0.32$. The field is averaged over  the azimuthal angle and 
approximately $2\times 10^6$ collisions).  A  torroidal
convection roll  is clearly present   in which the particles flow,  on
average, up from the center and down the side wall.  Note that, unlike
the related  Rayleigh-B{\'e}nard phenomenon,  the roll  is  asymmetric. The
velocity field  is in near quantitative  agreement  with that observed
experimentally by  Wildman et  al.  \cite{WildmanHP01}. In particular,
the center of the vortex roll is at $(r/ \sigma \simeq 10, z/ \sigma \simeq 6)$ in reduced
units  which corresponds to $(r\simeq    50{\rm  mm}, z\simeq  30{\rm mm})$   in
experiments\cite{WildmanHP01}.  The  shape  of the roll,   as well the
location of   its  center, do not   change  appreciably as  the shaker
amplitude is varied between $0.28\leq A/\sigma\leq 0.37$.

\begin{figure}
\resizebox{7.5cm}{!}{\includegraphics{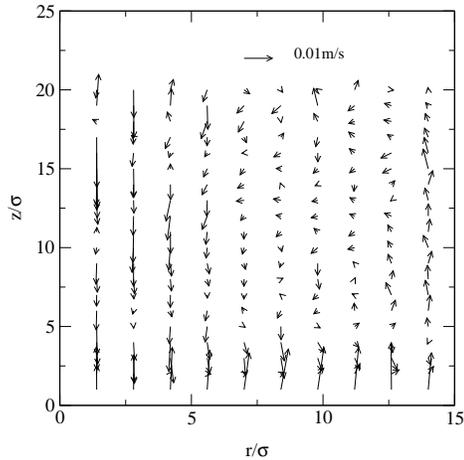}}
\caption{Same as Figure \ref{fig:1} except that the wall-particle 
 restitution coefficient is
$c_w=1.0$. Note that the velocities are much smaller than in 
Figure \ref{fig:1}. }\label{fig:2}
\end{figure}

To  highlight the role  played by  the wall,  we performed  additional
simulations in which the wall-sphere collision is  taken as elastic (a
``perfect' wall for which the wall-particle restitution coefficient is
equal  to 1) with the  other parameters  unchanged.  The mean velocity
field  of  this simulation   is quite  different  from  the inelastic
case. Although  some   convection is apparent in   Figure \ref{fig:2},
(particles on average flow up the wall and  down to the center). It is
weaker  by about an order  of  magnitude in intensity  (compare the velocity
units at the    top  of Figs.~\ref{fig:1} and.~\ref{fig:2}) and the roll direction 
is {\em opposite}.

\begin{figure}
\resizebox{8cm}{!}{\includegraphics{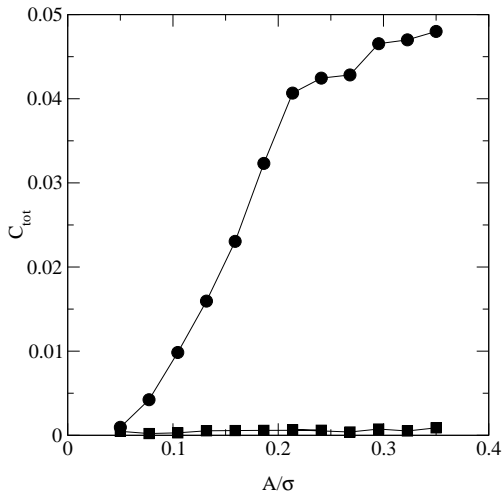}}
\caption{Order parameter $C_{\rm tot}$
versus the shaker amplitude $A/ \sigma $ of the  base wall. The squares and
circles correspond  to simulations where the wall-particle restitution
coefficient is $c_w=1.0$ and $0.68$, respectively .}\label{fig:6}
\end{figure}

To quantify the bulk  convection,  we  have calculated the   total
velocity  correlation, $C_{\rm  tot}=(1/N_c)\Sigma_{i,j}  C(i,j)$\cite{RRC00}
where $(i,j)$ label the cells in the hydrodynamic region, $N_c$ their
total number and
\begin{equation}
C(i,j)=(1/8)\Sigma_{i',j'}{\bf v}(i,j).{\bf v}(i',j')
\end{equation}
where   the sum $(i',j')$ is  over  the $8$ neighboring  cells of $(i,j)$.
See Figure~\ref{fig:6}. 
Bulk  convection is present  when $C_{\rm  tot}$ is different
from zero. 

For  $c_w=0.68$, as   the shaker  amplitude  is   increased the  total
velocity correlation $C_{\rm tot}$   becomes different from zero  when
$A/\sigma >0.05$, increases  until $a/\sigma\simeq 0.2$ and  stays roughly constant for
larger amplitudes.  A similar behavior was observed in the experiments
of Wildman  et al.  \cite{WildmanHP01}. For  $c_w=1$, $C_{\rm tot}$ is
almost equal to zero for all amplitudes (Figure
\ref{fig:6}).   A close examination of the corresponding flow velocity
fields show  that the amplitude of  the circulation of particles up the
wall and down to the center exists for $A/\sigma >0.05$, a feature which is
not well handled by the calculation of the order parameter.

\begin{figure}
\resizebox{8cm}{!}{\includegraphics{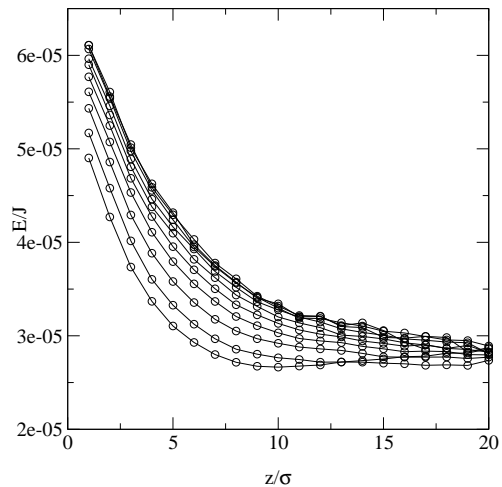}}
\caption{Granular temperature for a  vibrofluidized system of inelastic hard
spheres as a function of the altitude $z$  for different values of the
radial  position. From bottom to top,   the curves correspond to $r/ \sigma
=1.3,2.6,3.9,5.2,6.5,7.8,9.1,10.4,11.7,13  $.   The particle mass is
$1.875\times 10^{-4}kg$. }\label{fig:3}
\end{figure}

\begin{figure}
\resizebox{8cm}{!}{\includegraphics{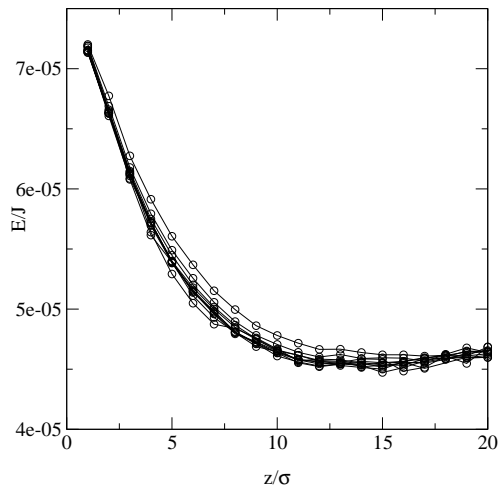}}
\caption{Same as Figure \ref{fig:3} except $c_w=1.0$. From top to bottom,   the curves correspond to $r/ \sigma
=13, 11.7, 10.4, 9.1, 7.8 6.5, 5.2, 3.9, 2.6, 1.3$.}
\label{fig:4}
\end{figure}

Figure~\ref{fig:3} shows the granular temperature as a function of the
altitude, $z$, for different values the radial position. These results
agree  well  with  the experiments   (Fig.1 of  Ref\cite{WildmanHP01})
except for  an underestimated  maximum  close to  the basewall.  These
differences  could result from the  base-particle collisions which are
assumed elastic in simulations.  A minimum in the granular temperature
profile is   observed   in the  current   simulations  (and   also  in
experiments) in  the neighborhood of  the center of  the cylinder, but
becomes weaker  close to the  wall cylinder.  Fig~\ref{fig:4} displays
the same profiles for the system with $c_w=1$: The temperature minimum
in the vertical   direction is more  pronounced, but   the temperature
varies little in the radial direction.  The existence  of a minimum in
the granular temperature along the $z$-direction is a phenomenon which
is   also present in  the   absence  of convection\cite{BreyRM01}  and
results from the fact that at high altitudes the density and collision
frequency are small, leading to a small number of hot particles.

\begin{figure}
\resizebox{7cm}{!}{\includegraphics{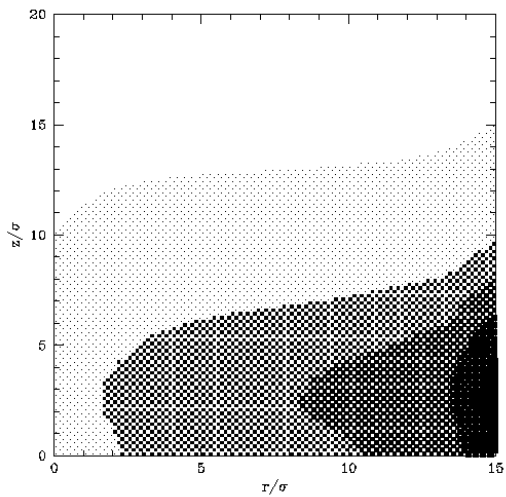}}
\caption{Packing fraction ($\eta$) plot for the  vibrofluidized system of
inelastic  hard    spheres,  with $c_w=0.68$.   The    greyscale  step
corresponds to an  $\eta$ increase  of $0.02$ and  white
regions   correspond  to $\eta<0.02$.}\label{fig:5}
\end{figure}

Figure~\ref{fig:5} displays  a packing fraction plot which agrees qualitatively
with the experimental results, except in the bottom region of the side
wall where the  density is higher than  in experiments.  Nevertheless,
there is a maximum in  the density profile  in the vertical direction,
whatever the radial distance.  Moreover, there is also a maximum along
the radial direction which is  a consequence both of the  compressible
nature  of   the inelastic  hard   sphere   system  and of  the  small
wall-particle  restitution coefficient. When  $c_w=1$, 
the radial density gradient almost vanishes.

Although a complete theoretical treatment of  the convection is
beyond the scope of this Letter, we present some simple arguments that
may help  to  rationalize the observed behavior.   The Rayleigh-B{\'e}nard
convection  which is the paradigm  for  pattern formation in classical
fluid  mechanics, has been extensively  studied   over the past  three
decades\cite{BPA00,DBB00}.  A  temperature gradient applied to a fluid
leads to a competition between buoyancy and disspation, which leads to
the occurence  of  rolls above  a threshold  characterized by  the non
dimensional Rayleigh number  $R_a$,  which characterizes the  ratio of
the two forces. For    larger  temperature gradients, on   observes  a
cascade of bifurcations ending in a chaotic regime.

For classical  fluids, hydrodynamic  equations reproduce the  observed
patterns remarkably well. The range of validity of continuum equations
applied to granular materials, which contain a sink term in the energy
equation  to account for the  dissipation of energy during collisions,
is more    restricted.   Hydrodynamic equations   can  provide  a good
description of  inelastic   spheres  \cite{D01},   provided that   the
particles packing fraction  remains $\eta >0.04$  (below which  the mean free path
becomes comparable to the diameter of the cylinder). Similarly, if the
packing fraction is too large ($\eta>0.3$) the dissipation  becomes too strong and
the  description   breaks  down.   Using such  an   approach,  Brey et
al.\cite{BreyRM01} obtained temperature profiles  for an open granular
system,  {\em  without convection}   and predicted the  presence of  a
minimum as a function of $z$, coinciding with the limit of validity of
the hydrodynamic equations.

When $c_w=0.68$, the hydrodynamic description breaks down close to the
wall (high density and  temperature gradient, strong dissipation). The
onset of the  bulk convective regime with a  ``real'' wall is  lowered
because  of the  existence  of the    radial  temperature gradient  in
addition to the vertical one.  This can explain  both the direction of
circulation and the lowering  of the threshold of  convection compared
to  the  case when   the wall is  perfect  and  no bulk  convection is
present.   The circulation of particles  is  restricted  close to  the
sidewall     where  strong dissipation  is   not    well described  by
hydrodynamic equations.  A kinetic  approach is required close to  the
cylinder.
 
Our simulations  clearly demonstrate the   relevance of the  inelastic
hard sphere model for describing the  experimental study of Wildman et
al\cite{WildmanHP01} and underline the key  influence of the  cylinder
wall on the  convection.  Experimentally,  unlike in the  simulations,
the wall-particle restitution  coefficient  is not an  easily  tunable
parameter.  We therefore propose a modification to the experiment that
may   lead to the   observation  of new  phenomena.   Specifically, we
suggest adding an  inner (solid) cylinder  of a given radius,  coaxial
with  the original  cylinder.    In the  neighborhood of   this  inner
cylinder, we expect particles  to have a  downward motion, and two, or
more, toroidal rolls with alternate directions of circulation, as in a
fluid system, might be produced.

We thank R.  Wildman for supplying his  experimental data, E.  Cl{\'e}ment
P.   Manneville,  and  R.  Ramirez    for fruitful  discussions.  J.T.
acknowledges support from the National Science Foundation (CHE-9814236) 
and the Centre National de la R{\'e}cherche Scientifique.


\begin{thebibliography}{10}

\bibitem{WildmanHP01}
R.~D. Wildman, J.~M. Huntley, and D.~J. Parker, Phys. Rev. Lett. {\bf 86},
  3304  (2001).

\bibitem{EvesqueR89}
P. Evesque and J. Rajchenbach, Phys. Rev. Lett. {\bf 62},  44  (1989).

\bibitem{WildmanHHPA00}
R.~D. Wildman, J.~M. Huntley, J.~P. Hansen, D.~J. Parker, and D.~A. Allen,
  Phys. Rev. E {\bf 62},  3826  (2000).

\bibitem{RRC00}
R. Ramirez, D. Risso, and P. Cordero, Phys. Rev. Lett. {\bf 85},  1230  (2000).

\bibitem{G97}
E.L. Grossman, Phys. Rev. E {\bf 56},  3290  (1997).

\bibitem{van-NoijeE00}
T.~P.~C. van Noije and M.~H. Ernst, Phys. Rev. E {\bf 61},  1765  (2000).

\bibitem{van-NoijeETP99}
T.~P.~C. van Noije, M.~H. Ernst, E. Trizac, and I. Pagonabarraga, Phys. Rev. E
  {\bf 59},  4326  (1999).

\bibitem{Javier-BreyDCS98}
J. J. Brey. J., J.W. Dufty , C.S. Kim, and A. Santos.  Phys. Rev. E {\bf 58},  4638
  (1998).

\bibitem{WHP01}
R.~D. Wildman, J.~M. Huntley, and D.~J. Parker, Phys. Rev. E {\bf 63},  061311
  (2001).

\bibitem{MB97}
S. McNamara and J.-L. Barrat, Phys. Rev. E {\bf 55},  7767  (1997).

\bibitem{DeltourB97}
P. Deltrour and J.-L. Barrat, J. Phys. I France {\bf 7},  137  (1997).

\bibitem{BreyRM01}
J.~J. Brey, M.~J. Ruiz-Montero, and F. Moreno, Phys. Rev. E {\bf 63},  061305
  (2001).

\bibitem{BPA00}
E. Bodenschatz, W. Pesch, and G. Ahlers, Annu.. Rev. Fluid. Mech. {\bf 32},
  709  (2000).

\bibitem{DBB00}
K. E. Daniels, B. B. Plapp, and E. Bodenschatz, Phys. Rev. Lett. {\bf 84},  5320
  (2000).

\bibitem{D01}
J. Dufty, Condmat/0109215  (2001).

\end{thebibliography}
\end{document}